\documentstyle[12pt]{article}
\def\LP{\displaystyle{\Biggl(}}
\def\RP{\displaystyle{\Biggr)}}
\newcommand{\eq}{\begin{equation}}
\newcommand{\intx}{\int d^4 \! x \, }
\begin{document}
\baselineskip22pt
\def\ftoday{{\sl  \number\day \space\ifcase\month
\or Janvier\or F\'evrier\or Mars\or avril\or Mai
\or Juin\or Juillet\or Ao\^ut\or Septembre\or Octobre
\or Novembre \or D\'ecembre\fi
\space  \number\year}}

\titlepage
\begin{center}
{\huge  Antisymmetric Tensor Gauge Theory}

\vspace{.5cm}
{\Large G.Bandelloni}
\vspace{.5cm}

{\it
Istituto di Fisica della Facolt\'a di Ingegneria -- Universit\`a di Genova\\
Istituto Nazionale di Fisica Nucleare -- sez. di Genova\\
Via Dodecaneso, 33 -- 16146 Genova (Italy)}

\vspace{.5cm}
and

\vspace{.4cm}
{\Large A.Blasi}

\vspace{.5cm}
{\it
Dipartimento di Fisica  -- Universit\`a di Genova\\
Istituto Nazionale di Fisica Nucleare -- sez. di Genova\\
Via Dodecaneso, 33 -- 16146 Genova (Italy)}

\end{center}

\vspace{1cm}

\begin{center}
\bf ABSTRACT
\end{center}
{\it
We analyze to all perturbative orders the properties of two possible
quantum extensions of classically on-shell equivalent antisymmetric tensor
gauge models in four dimensions. The first case, related to the soft
 breaking of a topological theory wants a gauge field of canonical dimension
one. The other possibility, which assigns canonical dimension two to the gauge
 field, leads to the $\sigma$ model interpretation of the theory. In both
 instances we find that the models are anomaly free.}

\vfill
GEF-Th-10/93 \\
\hfill March 1993
\newpage

\section{Introduction}

The antisymmetric tensor field models have been introduced several years ago
\cite{1}; at the time the interest was focused on their dynamical
relation to four dimensional $\sigma$ models and string theories
\cite{2,3,4} and only later it was realized
that, in a different regime, they describe a topological theory which can be
viewed as the natural generalization of the three-dimensional Chern-Simons
model \cite{5}. These different aspects can be illustrated, at the classical
level,
by the action

\begin{equation}
 I =- {1 \over 2}
  \intx \varepsilon^{\mu\nu\rho\sigma}
                     F^{a}_{\mu\nu} B^{a}_{\rho\sigma}
          -{k\over 2}\intx{\ } A^{a}_{\mu} A^{a\mu}
-{l \over 4}\intx F^{a}_{\mu\nu}F^{a\mu\nu}
\label{11}
\end{equation}

where

\begin{equation}
 F^{a}_{\mu\nu} = \partial_{\mu} A^{a}_{\nu} -
                    \partial_{\nu} A^{a}_{\mu} +
                    f^{abc} A^{b}_{\mu}A^{c}_{\nu}
\label{12}
\end{equation}

and $B^{a}_{\rho\sigma}=-B^{a}_{\sigma\rho}$ is the antisymmetric tensor field.

$I$ is invariant under the B.R.S.gauge transformations

\begin{eqnarray}
\delta  A^{a}_{\mu}{\ } &=& 0 \nonumber \\
\delta  B^{a}_{\mu\nu} &=&
{({\ }D_{\mu} c_{\nu} - D_{\nu} c_{\mu} {\ })}^{a}
\label{13}
\end{eqnarray}

 with

\begin{equation}
 D_{\mu}^{ab} = \partial_{\mu} \delta^{ab} + f^{acb} A_{\mu}^{c}
\label{14}
\end{equation}
\noindent

and yields the equations of motion
\begin{eqnarray}
     { \delta I \over \delta B^{a}_{\mu\nu}} &=& - {1 \over 2}
  \varepsilon^{\mu\nu\rho\sigma}
                     F^{a}_{\mu\nu} = 0        \nonumber     \\
{\delta I \over \delta A^{a}_{\mu}} &=& \varepsilon^{\mu\nu\rho\sigma}
{(D_{\nu} B_{\rho\sigma})}^{a} - kA^{a\mu} + l(D_{\nu}F^{\mu\nu})^{a} = 0
\label{15}
\end{eqnarray}

which imply

\begin{eqnarray}
A^{a}_{\mu} &=& (U^{-1}\partial_{\mu}U)^{a} \nonumber \\
 k{(D_{\mu} A^{\mu})}^{a} &=& 0
\label{16}
\end{eqnarray}

In a suitable parametrization by means of scalar fields for the $U$ matrix,
(\ref{16}) yields the characteristic $\sigma$ model constraint.
Proceeding in this direction we are led to attribute to the gauge field a
canonical dimension equal to two \cite{4} and therefore the $l$ coupling should
be
set to zero. Within this choice the symmetry (\ref{13}) and the power
counting are sufficient to identify the classical action.

A different interpretation is provided by the topological regime $k = l = 0$
where the classical action does not contain the metric tensor. Here the
switching on of the $k$ coupling can be seen as the soft breaking
\footnote{The softness of the breaking refers here only to the power
counting; indeed the complete classical action, including the gauge fixing,
of the true topological model possesses an additional vectorial
supersymmetry \cite{6} which is definitely broken by the $k$ coupling}of an
 otherwise topological theory if the gauge field is assigned canonical
dimension equal to one. In this case the symmetry (\ref{13})
offers no protection against the appearance of higher dimensionality
 $A^{a}_{\mu}$ couplings and therefore we enlarge the invariance to include
a gauge transformation for $A_{\mu}^{a}$ and consider the model for
nonvanishing $l$.

The two approaches which classically describe  on-shell identical models, are
expected to generate different quantum extensions.
The interest in considering them is twofold: first we shall provide a
complete characterization of these extensions by analyzing the cohomology
spaces of the corresponding nilpotent B.R.S. operators.
In a perturbative framework and for both cases,  we
exclude the presence of anomalies, a result which was only known for the
topological BF model \cite{7}
Second, for nonvanishing $k$ and $l$ we find that the propagators are
analytic in these parameters and therefore the model should go smoothly
to the topological one which is known to be completely finite.
In this case we find that the finiteness property does not hold since there
are two renormalization constants, but the counterterms are still given
by a B.R.S. cocycle.

The difference between these two ways of looking at the same classical problem
becomes apparent at the quantum level since
 the quantum extensions under consideration are defined
by different symmetries and therefore do not coincide off-shell and in
the gauge fixing sectors. We have unified notations where possible and have
chosen to present each approach separetely. Thus the Section "Quantum
 extension 1" contains the definitions, the strategy and the results
concerning the softly broken model with $A^{a}_{\mu}$ of dimensionality one,
 while the same
 arguments for the case with $A^{a}_{\mu}$ of dimensionality two
 reported in the Section
"Quantum extension 2" .The technical aspects of the treatment are collected
in the Appendices, while comments on both extensions are contained in the
conclusive Section.

\section{Quantum extension 1}

We start with the classical action

\begin{equation}
 I_{0} =- {1 \over 2}
 \intx{\ } \LP \varepsilon^{\mu\nu\rho\sigma}
F^{a}_{\mu\nu} B^{a}_{\rho\sigma} +kA^{a}_{\mu} A^{a\mu}
+{l\over 2} F^{a\mu\nu} F^{a}_{\mu\nu}\RP
\label{21}
\end{equation}

which is invariant (at $k=0$) under (\ref{13}) and the gauge transformations
\begin{eqnarray}
{\delta^{(1)}}A^{a}_{\mu} &=& {(D_{\mu}\theta)}^{a} \nonumber \\
 {\delta^{(1)}}B^{a}_{\mu\nu} &=& f^{abc} B^{b}_{\mu\nu}\theta^{c}
\label{22}
\end{eqnarray}
The $k$ coupling breaks softly the $\delta^{(1)}$ symmetry since
\begin{equation}
 \delta^{(1)} I_{0} = k\int d^4x{\ } \theta^{a} \partial_{\mu} A^{a\mu}
\label{23}
\end{equation}

This signals the fact that if we gauge fix the $\delta^{(1)}$
  transformation
in the Landau gauge, we could be able to reabsorb the breaking (\ref{23})
by means of the corresponding Lagrange multiplier field, which, being free,
allows a complete control of the quantum corrections.

The gauge fixing of the invariances (\ref{13};\ref{22}) is by now well known
\cite{3,8}; the corresponding non-vanishing BRS transformations are

\begin{eqnarray}
         s A^{a}_{\mu}{\ } &=&  {(D_{\mu}\theta)}^{a} \nonumber   \\
       s \theta^{a}{\ }{\ } &=& {1 \over 2}f^{abc}\theta^b \theta^c \nonumber
\\
         s c^{a}_{\mu}{\ }{\ } &=&  {(D_{\mu}d)}^{a} +
                         f^{abc}\theta^{c}c_{\mu}^{b}  \nonumber  \\
         s d^{a}{\ }{\ }   &=& f^{abc}\theta^{b}d^{c}      \nonumber \\
         s B^{a}_{\mu\nu} &=&  {(D_{\mu}c_{\nu} - D_{\nu}c_{\mu})}^{a}
                 + f^{abc}B^{b}_{\mu\nu}\theta^{c}
                 + f^{abc}\varepsilon_{\mu\nu\rho\sigma}
                   (\partial^{\rho}{\bar c}^{b\sigma})d^{c}   \nonumber \\
         s {\bar c}^{a}_{\mu}{\ }{\ } &=& b^{a}_{\mu}\nonumber \\
         s {\bar \theta}^{a}{\ }{\ } &=& b^{a}{\ }\nonumber \\
         s {\overline{\overline{d}}}^{a}{\ }{\ } &=& {\bar b}^{a} \nonumber \\
         s e^{a}{\ }{\ } &=& \lambda^{a}
\label{24}
\end{eqnarray}

The gauge fixing action is given by

\begin{eqnarray}
   I_{gf} &=&  \intx \LP b^{a}\partial_{\mu} A^{a\mu}
- {\bar \theta}^{a} \partial^{\mu}{(D_{\mu}\theta)}^{a} +
       b^{a\mu}\partial^{\nu}B^{a}_{\mu\nu}\nonumber \\
 &-& {l \over 8}b^{a\mu}((\partial_{\rho}
       \partial^{\rho} - {k \over l})\delta^{\nu}_{\mu} -
\partial_{\mu}\partial^{\nu})
        b^{a}_{\nu} +
       b^{a}_{\mu}(\partial^{\mu}e^{a})\nonumber  \\
       &+& (\partial_{\mu} {\bar c}^{a\mu})\lambda^{a} -
       (\partial^{\mu}{\overline{\overline{d}}}^{a})
       ({\ }{(D_{\mu}d)^{a}}  + f^{abc}\theta^{c}c_{\mu}^{b}{\ })\nonumber \\
     &+& (\partial^{\nu} {\bar c}^{a\mu})
         ({\ }{(D_{\mu}c_{\nu})^{a}} - {(D_{\nu}c_{\mu})^{a}} +
          f^{abc}B^{b}_{\mu\nu}\theta^{c} {\ })\nonumber \\
     &+&{1 \over 2}f^{abc} \varepsilon^{\mu\nu\rho\sigma}
         (\partial_{\mu}{\bar c}^{a}_{\nu})
         (\partial_{\rho}{\bar c}^{b}_{\sigma})d^{c} {\ }{\ }
         + {\bar b}^{a}\partial^{\mu}c_{\mu}^{a}\RP ,
\label{25}
\end{eqnarray}

and the external field contribution is

\begin{eqnarray}
   I_{ef} &=&   \intx \LP \gamma^{a\mu\nu}( s B^{a}_{\mu\nu} ) +
        \gamma^{a\mu}( s A^{a}_{\mu} ) + \xi^{a}( s \theta^{a} ) \nonumber \\
    &+&  \eta^{a}( s d^{a} )
       + \Omega^{a\mu} ( s c^{a}_{\mu})  +
     {1 \over 2}{\ }f^{abc} \varepsilon_{\mu\nu\rho\sigma}
     \gamma^{a\mu\nu}\gamma^{b\rho\sigma}d^{c} \RP ,
\label{26}
\end{eqnarray}

The fact that the $s$ transformation is nilpotent only on-shell, since

\begin{equation}
 s^{2}B^{a}_{\mu\nu} = - f^{abc}F^{b}_{\mu\nu}d^{c}
\label{27}
\end{equation}

is accounted for, in (\ref{26}), by the bilinear $\gamma^{a}_{\mu\nu}$
coupling.

The quantum number assignement for the quantized fields is

\begin{center}
\begin{tabular}{|l|r|r|r|r|r|r|r|r|r|r|r|r|r|}\hline
&$A_{\mu}$ & $B_{\mu\nu}$ & $\theta$ & $\bar{\theta}$&$b$& $c_{\mu}$ &
$\bar c_{\mu}$& $b_{\mu}$ & $d$ & $\overline{\overline{d}}$ &
$\bar b$ & $e$ & $\lambda$ \\ \hline
dim&1&2&0&2&2&1&1&1&0&2&2&2&2\\ \hline
$\Phi\Pi$&0&0&1&-1&0&1&-1&0&2&-2&-1&0&1\\ \hline
G-parity&+&-&+&+&+&-&-&-&-&-&-&-&- \\ \hline
\end{tabular}
\end{center}
and for the external ones is

\begin{center}
\begin{tabular}{|l|r|r|r|r|r|}\hline
&$\gamma_{\mu}$ & $\xi$ & $\eta$ & $\gamma_{\mu\nu}$ & $\Omega_{\mu}$ \\ \hline
dim&3&4&4&2&3 \\ \hline
$\Phi\Pi$&-1&-2&-3&-1&-2 \\ \hline
G-parity&+&+&-&-&- \\ \hline
\end{tabular}
\end{center}

\noindent
The classical action
\begin{equation}
I^{cl} = I_{0} + I_{gf} + I_{ef}
\label{icl}
\end{equation}

\noindent
is G-parity even, $\Phi\Pi$ neutral and  satisfies the softly broken BRS
identity
\begin{eqnarray}
\intx \LP {\delta I^{cl} \over {\delta{\gamma_{\mu}^{a}}}}{\delta I^{cl}
\over {\delta{A^{a\mu}}}}
   &+& {\delta I^{cl} \over {\delta{\gamma_{\mu\nu}}}}
{\delta I^{cl} \over {\delta{B^{a\mu\nu}}}} +
   {\delta I^{cl} \over {\delta{\xi^{a}}}}{\delta I^{cl} \over
{\delta{\theta^{a}}}} +
  {\delta I^{cl} \over {\delta{d^{a}}}}{\delta I^{cl} \over {\delta{\eta^{a}}}}
\nonumber \\
 + {\delta I^{cl} \over {\delta{c_{\mu}^{a}}}}
{\delta I^{cl} \over {\delta{\Omega^{a\mu}}}}
&+& b^{a}{\delta I^{cl} \over {\delta{\bar \theta^{a}}}} +
 b_{\mu}^{a}{\delta I^{cl} \over {\delta{\bar c_{\mu}^{a}}}} +
{\bar b^{a}}{\delta I^{cl} \over {\delta{\overline{\overline{d^{a}}}}}} +
{\lambda^{a}}{\delta I^{cl}\over{\delta{e^{a}}}}\RP \nonumber \\
 &=& k{\intx \theta^{a}\partial_{\mu}A^{a\mu}}
\label{28}
\end{eqnarray}

\noindent
the local equation of motion symmetries
\begin{eqnarray}
T^{a}(x)I^{cl}&\equiv& {\delta I^{cl} \over {\delta b^{a}(x)}} =
 \partial_{\mu}A^{a\mu}(x) \nonumber \\
{\bar T^{a}}(x)I^{cl}&\equiv& {\delta I^{cl} \over {\delta
{\bar\theta}^{a}(x)}} +
{\partial_{\mu}}{\delta I^{cl} \over {\delta \gamma^{a}_{\mu}(x)}} = 0
\nonumber \\
R^{a}_{\mu}(x)I^{cl}&\equiv& {\delta I^{cl} \over {\delta {b_{\mu}}^{a}(x)}
} = {\partial^{\nu}{B^{a}_{\mu\nu}}(x)} -{l \over 4}{(({\partial_{\rho}}
{\partial^{\rho}} - {k \over l})\delta_{\mu\nu} - \partial_{\mu}\partial^{\nu})
b^{a}_{\nu}(x)} + {\partial_{\mu}e^{a}(x)} \nonumber \\
{\bar R^{a}_{\mu}}(x)I^{cl}&\equiv& {\delta I^{cl} \over {\delta {\overline c}
^{a}_{\mu}(x)}} +{\partial_{\nu}}{\delta I^{cl} \over {\delta {\gamma^{a}
_{\mu\nu}}^{a}(x)}} = - {\partial_{\mu}\lambda^{a}}(x)  \nonumber \\
U^{a}(x)I^{cl}&\equiv& {\delta I^{cl} \over {\delta {\bar b}^{a}(x)}} =
 {\partial_{\mu}c^{a\mu}}\nonumber \\
{\bar U^{a}}(x)I^{cl}&\equiv& {\delta I^{cl} \over {\delta {\overline
{\overline d}}^{a}(x)}} - {\partial_{\mu}}{\delta I^{cl} \over {\delta
{\Omega}^{a\mu}(x)}} = 0 \nonumber \\
{\overline V}^{a}(x)I^{cl}&\equiv& {\delta I^{cl} \over
{\delta {e}^{a}(x)}} = {\partial_{\mu} b^{a\mu}}(x) \nonumber \\
V^{a}(x)I^{cl}&\equiv& {\delta I^{cl} \over {\delta {\lambda}^{a}(x)}}
 ={\partial_{\mu}{\bar c}^{a\mu}}(x)
\label{29}
\end{eqnarray}

and the integrated $d^{a}$ ghost equation of motion \cite{9}
\begin{eqnarray}
N^{a}I^{cl}&\equiv& \intx ({\delta I^{cl} \over {\delta {d}^{a}}} =
-\intx f^{abc}{\overline{\overline d}^{b}}{\delta I^{cl} \over
{\delta {b}^{c}}}) \nonumber \\
 &+& f^{abc}\intx (\varepsilon^{\mu\nu\rho\sigma}\gamma_{\mu\nu}^{b}(
\partial_{\rho}{\bar c}^{c}_{\sigma} + {1 \over 2}\gamma^{c}_{\rho\sigma}) +
\eta^{b}\theta^{c} + \Omega^{c\mu}A^{b}_{\mu})
\label{210}
\end{eqnarray}

\noindent
The defining symmetries (\ref{28};\ref{29};\ref{210}) are the starting point to
analyze
the perturbative renormalizability of the model; setting
\begin{equation}
\Gamma = I^{cl} +\sum_{n=1} \Gamma^{(n)}\hbar^{n}
\label{211}
\end{equation}

we have that $\Gamma^{(n)}$ obeys
\begin{eqnarray}
S_{L}\Gamma^{(n)}\equiv
\intx \LP {\delta I^{cl} \over {\delta{\gamma_{\mu}^{a}}}}
{\delta \Gamma^{(n)} \over {\delta{A^{a\mu}}}}
&+& {\delta \Gamma^{(n)} \over {\delta{\gamma_{\mu}^{a}}}}
{\delta I^{cl} \over {\delta{A^{a\mu}}}}
+ {\delta I^{cl} \over {\delta{\gamma_{\mu\nu}}}}
{\delta \Gamma^{(n)} \over {\delta{B^{a\mu\nu}}}} +\nonumber \\
{\delta \Gamma^{(n)} \over {\delta{\gamma_{\mu\nu}}}}
{\delta I^{cl} \over {\delta{B^{a\mu\nu}}}}
&+& {\delta I^{cl} \over {\delta{\xi^{a}}}}
{\delta \Gamma^{(n)} \over {\delta{\theta^{a}}}} +
{\delta \Gamma^{(n)} \over {\delta{\xi^{a}}}}
{\delta I^{cl} \over {\delta{\theta^{a}}}} +
{\delta I^{cl} \over {\delta{d^{a}}}}{\delta \Gamma^{(n)}
\over {\delta{\eta^{a}}}} \nonumber \\
+ {\delta \Gamma^{(n)} \over {\delta{d^{a}}}}
{\delta I^{cl} \over {\delta{\eta^{a}}}}
&+& {\delta I^{cl} \over {\delta{c_{\mu}^{a}}}}
{\delta \Gamma^{(n)} \over {\delta{\Omega^{a\mu}}}}
 + {\delta \Gamma^{(n)} \over {\delta{c_{\mu}^{a}}}}
{\delta I^{cl} \over {\delta{\Omega^{a\mu}}}}
 b^{a}{\delta \Gamma^{(n)} \over {\delta{\bar \theta^{a}}}} +
 b_{\mu}^{a}{\delta \Gamma^{(n)} \over {\delta{\bar c_{\mu}^{a}}}} + \nonumber
\\
{\bar b^{a}}{\delta \Gamma^{(n)} \over {\delta{\overline{\overline{d^{a}}}}}}
&+& {\lambda^{a}}{\delta \Gamma^{(n)}\over{\delta{e^{a}}}}
 - k \theta^{a}{\delta \Gamma^{(n)} \over b^{a}}\RP
\label{212}
\end{eqnarray}
\begin{eqnarray}
T^{a}(x)\Gamma^{(n)} &=& 0  \nonumber \\
{\bar T^{a}}(x)\Gamma^{(n)} &=& 0 \nonumber \\
R^{a}_{\mu}(x)\Gamma^{(n)} &=& 0 \nonumber \\
{\bar R^{a}_{\mu}}(x)\Gamma^{(n)} &=& 0  \nonumber \\
U^{a}(x)\Gamma^{(n)} &=& 0 \nonumber \\
{\bar U^{a}}(x)\Gamma^{(n)} &=& 0 \nonumber \\
{\overline V}^{a}(x)\Gamma^{(n)} &=& 0  \nonumber \\
V^{a}(x)\Gamma^{(n)} &=& 0 \nonumber \\
N^{a}\Gamma^{(n)} &=& 0
\label{213}
\end{eqnarray}

\noindent
The linearized operator $S_{L}$ in (\ref{212}) is not nilpotent due to the
$k$ coupling; by direct computation we find:
\begin{eqnarray}
S_{L}^{2}\equiv -k\intx ( {\partial_{\mu}\theta^{a}}{\delta \over
{\delta {\gamma^{a}_{\mu}}}} &+&{\partial_{\mu}A^{a\mu}}{\delta \over
{\delta {\xi}^{a}}}) + \nonumber \\
-k\intx (\theta^{a}({\delta \over
{\delta {\overline \theta}^{a}}} &+& {\partial_{\mu}}{\delta \over
{\delta {\gamma}^{a}_{\mu}}})) -{1 \over 2}kf^{abc}\intx
{\theta^{b}}{\theta^{c}}
{\delta \over {\delta {b}^{a}}}
\label{214}
\end{eqnarray}

\noindent
and therefore on $\Gamma^{(n)}$ the nilpotency condition reduces to

\begin{equation}
D\Gamma^{(n)}\equiv \intx ( {\partial_{\mu}\theta^{a}}{\delta \over
{\delta {\gamma^{a}_{\mu}}}} +{\partial_{\mu}A^{a\mu}}{\delta \over
{\delta {\xi}^{a}}})\Gamma^{(n)} = 0
\label{215}
\end{equation}

\noindent
The other algebraic relations are
\begin{eqnarray}
{[S_{L}, T^{a}(x)]} &=& {\overline T}^{a}(x) \nonumber \\
{[S_{L}, R^{a}_{\mu}(x)]} &=& {\overline R}^{a}_{\mu}(x) \nonumber \\
{[S_{L}, U^{a}(x)]} &=& {\overline U}^{a}(x) \nonumber \\
{[S_{L}, V^{a}(x)]} &=& {\overline V}^{a}(x) \nonumber \\
{[S_{L}, D^{a}]} \equiv G^{a} &=&
-f^{abc}\intx ({\eta^{b}{\delta \over {\delta {\xi^{c}}}}} +
 {\Omega^{b}_{\mu}}{\delta \over {\delta\gamma^{c}_{\mu}}}) \nonumber \\
 &-& f^{abc}\intx ({1 \over
4}\varepsilon^{\mu\nu\rho\sigma}{\gamma^{c}_{\mu\nu}}
{\delta \over {\delta B^{b\rho\sigma}}} +{\theta^{b}}
{\delta \over {\delta d^{c}}} +{A^{b}_{\mu}}
{\delta \over {\delta c^{c}_{\mu}}})
\label{216}
\end{eqnarray}

The analysis of the quantum extension of the symmetries identifing
the model begins with the functional relations (\ref{213}) which
imply that $\Gamma^{(n)}$ is  independent of the fields
$b^{a}(x), b^{a}_{\mu}(x), e^{a}(x), \lambda^{a}(x), {\overline b}^{a}(x)$
and it  depends on the external fields $\gamma^{a}_{\mu},
\gamma^{a}_{\mu\nu}, \Omega^{a}_{\mu}$ and the antighosts ${\overline\theta}^
{a}(x), {\overline c}^{a}_{\mu}(x), {\overline{\overline d}}^{a}(x)$
 through the linear combinations

\begin{eqnarray}
{\widehat\gamma}^{a}_{\mu}(x) &=& \gamma^{a}_{\mu}(x) + \partial_{\mu}
{\overline\theta}^{a}(x) \nonumber \\
{\widehat\gamma}^{a}_{\mu\nu}(x) &=& \gamma^{a}_{\mu\nu}(x) +{1 \over 2}
{( \partial_{\mu}{\overline c}^{a}_{\nu}(x) - \partial_{\nu}
{\overline c}^{a}_{\mu}(x))} \nonumber \\
{\widehat\Omega}^{a}_{\mu}(x) &=& \Omega^{a}_{\mu}(x) - \partial_{\mu}
{\overline{\overline d}}^{a}(x)
\label{31}
\end{eqnarray}

\noindent
while  $d^{a}(x)$ appears only as $\partial_{\mu}d^{a}(x)$. Substituting
the relations (\ref{213}) into (\ref{212}) we obtain the identity
\begin{eqnarray}
{\widehat S_{L}}\Gamma^{(n)}\equiv
\intx \LP {\delta {\widehat I} \over {\delta{{\widehat\gamma}_{\mu}^{a}}}}
{\delta \Gamma^{(n)} \over {\delta{A^{a\mu}}}}
&+& {\delta \Gamma^{(n)} \over {\delta{{\widehat \gamma}_{\mu}^{a}}}}
{\delta {\widehat I} \over {\delta{A^{a\mu}}}}
+ {\delta{\widehat I} \over {\delta{{\widehat \gamma}_{\mu\nu}}}}
{\delta \Gamma^{(n)} \over {\delta{B^{a\mu\nu}}}} \nonumber \\
+{\delta \Gamma^{(n)} \over {\delta{{\widehat \gamma}_{\mu\nu}}}}
{\delta {\widehat I} \over {\delta{B^{a\mu\nu}}}}
 &+& {\delta {\widehat I} \over {\delta{\xi^{a}}}}
{\delta \Gamma^{(n)} \over {\delta{\theta^{a}}}} +
{\delta \Gamma^{(n)} \over {\delta{\xi^{a}}}}
{\delta {\widehat I} \over {\delta{\theta^{a}}}} +
  {\delta {\widehat I} \over {\delta{d^{a}}}}
{\delta \Gamma^{(n)} \over {\delta{\eta^{a}}}} \nonumber \\
+ {\delta \Gamma^{(n)} \over {\delta{d^{a}}}}
{\delta {\widehat I} \over {\delta{\eta^{a}}}}
&+& {\delta {\widehat I} \over {\delta{c_{\mu}^{a}}}}
{\delta \Gamma^{(n)} \over {\delta{{\widehat \Omega}^{a\mu}}}}
 + {\delta \Gamma^{(n)} \over {\delta{c_{\mu}^{a}}}}
{\delta {\widehat I} \over {\delta{{\widehat\Omega}^{a\mu}}}}\RP
\label{32}
\end{eqnarray}

\noindent
where
\begin{eqnarray}
{\widehat I} &=& - {1 \over 2}
  \intx{\ } \LP \varepsilon^{\mu\nu\rho\sigma}
                     F^{a}_{\mu\nu} B^{a}_{\rho\sigma}
          +kA^{a}_{\mu} A^{a\mu}
         +{1\over 2} F^{a\mu\nu} F^{a}_{\mu\nu}\RP \nonumber \\
 &+&\intx \LP {\widehat \gamma}^{a\mu\nu}{((D_{\mu}c_{\nu} -
D_{\nu}c_{\mu})}^{a}
  - f^{abc}B^{b}_{\mu\nu}\theta^{c}) +
        {\widehat \gamma}^{a\mu}{(D_{\mu}\theta)}^{a}
  + \xi^{a}{1 \over 2}f^{abc}\theta^b \theta^c \nonumber \\
       &+& \eta^{a}f^{abc}\theta^{b}d^{c}
    +{\widehat \Omega}^{a\mu}({(D_{\mu}d)}^{a} +
f^{abc}\theta^{b}c_{\mu}^{c}) +
     {1 \over 2}{\ }f^{abc} \varepsilon_{\mu\nu\rho\sigma}
     {\widehat \gamma}^{a\mu\nu}{\widehat \gamma}^{b\rho\sigma}d^{c} {\ }\RP
\label{33}
\end{eqnarray}

\noindent
In Appendix B we  discuss the renormalization of the nilpotency condition
\begin{equation}
{\widehat D}\Gamma^{(n)}\equiv \intx ( {\partial_{\mu}\theta^{a}}{\delta \over
{\delta {{\widehat\gamma}^{a}_{\mu}}}} +{\partial_{\mu}A^{a\mu}}{\delta \over
{\delta {\xi}^{a}}})\Gamma^{(n)} = 0
\label{34}
\end{equation}
\noindent

By restricting the space to functionals obeying (\ref{34}) we have a nilpotent
${\widehat S}_{L}$ operator and in order to analyze
 the stability and the absence of anomalies of the model
we have to identify the integrated cohomology spaces of
${\widehat S}_{L}$ in the sectors with $\Phi\Pi$ charge zero and one.

 The integrated cohomology is identified through the descent system
 \cite{10}
\begin{eqnarray}
{{\widehat S}_{L}}X^{4}_{p}(x) &=&
\partial^{\mu}(X_{\mu})^{3}_{p+1}(x) \nonumber  \\
{{\widehat S}_{L}}(X_{\mu})^{3}_{p+1}(x) &=&
 \partial^{\nu}(X_{[\mu\nu]})^{2}_{p+2}(x) \nonumber  \\
{{\widehat S}_{L}}(X_{[\mu\nu]})^{2}_{p+2}(x) &=&
 \partial^{\rho}(X_{[\mu\nu\rho]})^{1}_{p+3}(x) \nonumber  \\
{{\widehat S}_{L}}(X_{[\mu\nu\rho]})^{1}_{p+3}(x) &=&
 \partial^{\sigma}(X_{[\mu\nu\rho\sigma]})^{0}_{p+4}(x) \nonumber  \\
{{\widehat S}_{L}}(X_{[\mu\nu\rho\sigma]})^{0}_{p+4}(x) &=& 0
\label{310}
\end{eqnarray}

\noindent
where the upper index denotes the dimensionality and the lower one the
$\Phi\Pi$ charge,  hence we will be interested to the cases $p = 0, 1$.
We emphasize that the analysis is not to be performed in the space of forms
since the classical action does not belong to this space due to the presence
of the terms $\intx kA^{a}_{\mu} A^{a\mu}$ and
$\intx F^{a\mu\nu} F^{a}_{\mu\nu}$.

The detailed discussion of the local cohomology spaces is performed in
Appendix C,
the results can be summarized by saying that the local functions $X^{ch}(x)$
of canonical dimension $\leq 4$,  even G-parity, satisfying the constraint
(\ref{215}) and which are cocycles modulo coboundaries of
${\widehat S}_{L}$ are the
gauge invariant functions of the sole field $\theta^{a}$ without space-time
derivatives.

It is now easy to reconstruct the solution of the descent equations (\ref{310})
since in the case $p = 0, 1$ the $X^{ch}(x)$ terms are absent due to
power counting and Lorentz indices content; thus we conclude that
\begin{equation}
X^{4}_{0,1}(x) = {\widehat S}_{L}X^{4}_{-1,0}(x) +
\partial^{\mu}X^{3}_{\mu0,1}(x)
\label{45}
\end{equation}

\noindent
and therefore
\begin{equation}
 X^{4}_{0,1} = \intx X^{4}_{-1,0}(x) = {\widehat S}_{L}{\intx X^{4}_{-1,0}(x)}
\label{46}
\end{equation}

This result ,which implies the absence of anomalies, also tells us that all
counterterms of the model  are BRS variations of local terms.Their
number can be computed by parametrizing the general local functional
${\widehat \Gamma}^{(0)}$ of dimensions $\leq 4$, even G-parity, $\Phi\Pi$
charge
 $-1$ and obeying
\begin{eqnarray}
{\widehat D}{\widehat \Gamma}^{(0)} &=& 0\nonumber \\
G^{a}{\widehat \Gamma}^{(0)} &=& 0 \nonumber \\
{\widehat\Gamma}^{0}&\not=& {\widehat S}_{L}{\widehat{\widehat\Gamma}}^{0}
\label{47}
\end{eqnarray}

\noindent
We find that there are only two nonvanishing parameters in ${\widehat\Gamma}^
{0}$; these can be identified with the multiplicative
renormalization constants of the fields $A^{a}_{\mu}(x), \theta^{a}(x)$.

\bigskip
\bigskip

\section{Quantum Extension 2}

The classical  action

\begin{equation}
 \Gamma^{cl}=-{1 \over 2}\int d^4x [\epsilon_{\mu\nu\rho\sigma}B_{\mu\nu}^a(x)
F_{\rho\sigma}^a(x)+A_{\mu}^a(x)
A_{\mu}^a(x)]
\label{1.1}
\end{equation}
\noindent
is invariant under the transformations

\begin{eqnarray}
\delta B_{\mu\nu}^a(x) &=&((D_{\mu}c_{\nu}(x) -D_{\nu}c_{\mu}(x)
)^a+\theta (gf^{abc}F^b_{\mu\nu}(x)c^c(x)) \nonumber \\
\delta c_{\rho}^a(x)&=&(D_{\rho}c(x))^a \nonumber \\
\delta \theta&=&-1
\label{1.2}
\end{eqnarray}
\noindent
i.e. in terms of the operator
\begin{eqnarray}
 \delta_{B.R.S.}&=& \int d^4x \biggl[((D_{\mu}c_{\nu}(x) -D_{\nu}c_{\mu}(x)
)^a+\theta gf^{abc}F^b_{\mu\nu}(x)c^c(x))
 \frac{\delta}{\delta B_{\mu\nu}^a(x)} \nonumber \\
&+& (D_{\rho}c(x))^a\frac{\delta}{\delta c_{\rho}^a(x) }\biggr]
-\frac{\partial}{\partial \theta}
\label{1.3}
\end{eqnarray}
\noindent

 Remark that without the $\theta$ term  the B.R.S. operator is
nilpotent only on-shell, so we obtain an "open gauge algebra"
\cite{13} definition of the model; the introduction of this ghost
(as a parameter carrying a negative unit of $\Phi\Pi$ charge) must be
considered as a "trick" intended to insure
the off-shell extension of the nilpotency condition, while it is required not
to modify the dynamics described by the classical action\cite{14}; hence

\begin{eqnarray}
\delta_{B.R.S.} \Gamma^{Cl} = 0   \nonumber \\
\frac{\partial\Gamma^{Cl}}{\partial\theta} = 0 \nonumber \\
\delta_{B.R.S.}^2 = 0
\label{1.4}
\end{eqnarray}

To  gauge fix  the invariance (\ref{1.2}) in the presence of the $\theta$
parameter we introduce  the  gauge functions $G_{\mu}^a$,
 Lagrange multipliers $b_{\mu}^a$ and antighosts $\bar{c}_{\mu}^a,\bar{c}^a$

\begin{eqnarray}
G_{\mu}^a(x)&=&(D^{\nu}B_{\mu\nu}(x))^a +b_{\mu}^a (x) \nonumber \\
 \delta {\bar{c}}_{\mu}^a(x)&=&{G_{\mu}^a(x)} \nonumber \\
 \delta \bar{c}^a (x)&=&(D^{\mu}\bar{c}_{\mu}(x))^a
\label{1.5}
\end{eqnarray}
\noindent

and the gauge fixing action
\begin{equation}
 \Gamma^{Gauge}=\int d^4x (\frac{1}{2}G_{\mu}^a(x)G^{a\mu}(x)-\bar{c}_
{\mu}^a(x)
(D^{\mu}D^{\nu}c_{\nu}(x))^a+ \bar{c}^a(x)(D^{\nu}D_{\nu}
c(x))^a)
\label{1.6}
\end{equation}

The $\delta$ variation of the Lagrange multiplier  is defined as
\begin{equation}
\delta b_{\mu}^a (x)=-(D_{\mu}D^{\nu}c_{\nu}(x))^a-
(D^{\nu}D_{\mu}c_{\nu}(x)-D^{\nu}D_{\nu}c_{\mu}(x))^a+
(\theta D^{\nu}F_{\mu\nu}(x))^a
\label{1.7}
\end{equation}
\noindent

which yields
\begin{equation}
\delta G_{\mu}^a(x)=-(D_{\mu}D^{\nu}c_{\nu}(x))^a
\label{1.8}
\end{equation}

With this choice we have at the classical level the functional relations
\begin{eqnarray}
\frac{\delta\Gamma}{\delta\bar{c}_{\mu}^a(x)}&=& \delta G^{a\mu}(x) \nonumber
\\
\frac{\delta\Gamma}{\delta {b}_{\mu}^a(x)} &=& G^{a\mu}(x) \nonumber \\
{D_{\mu}}\frac{\delta\Gamma}
{\delta\bar{c}^a(x)}&=& \delta^2 {b}_{\mu}^a(x) =
(D_{\mu}D^{\nu}D_{\nu}c(x))^{a}
\label{1.9}
\end{eqnarray}
\noindent
which imply the nilpotency of the $\delta_{B.R.S.}$ operator.
In order to transfer this framework at the quantum level we should introduce
external field couplings for all the B.R.S. variations of the fields; it is
well known that the procedure also implements the relations (\ref{1.9})
and allows a complete control of the vertex functional as far as the
dependence on the gauge fixing fields is concerned.
Taking into account this considerable simplification  we add only the sources
needed to discuss the residual symmetry, hence we define
\begin{equation}
\Gamma^0=\Gamma^{Cl} +\int d^4x \biggl[\gamma^{a\mu\nu}(x) \delta
B_{\mu\nu}^a(x)+
\zeta^{a\rho} (x) \delta c_{\rho}^a (x)\biggr]
\label{1.10}
\end{equation}
\noindent

and the linearized, nilpotent B.R.S. operator we consider is
\begin{equation}
 \delta_{B.R.S.}= \int d^4x \biggl[\frac{\delta \Gamma^0}{\delta
\gamma^{a\mu\nu}(x) } \frac{\delta}{\delta
B_{\mu\nu}^a(x) } + \frac{\delta\Gamma^0}{\delta
B_{\mu\nu}^a(x) }
\frac{\delta}{\delta
\gamma^{a\mu\nu}(x) }\]
\[+ \frac{\delta\Gamma^0}{\delta
\zeta^{a\nu}(x) }\frac{\delta}{\delta c_{\nu}^a(x) } +
\frac{\delta\Gamma^0}{\delta
c_{\nu}^a(x) }\frac{\delta}{\delta \zeta^{a\nu}(x) }\biggr]
-\frac{\partial}{\partial \theta}
\label{1.11}
\end{equation}
If we factorize the $\theta$ dependent part, we can write it as:
\begin{equation}
\delta_{B.R.S.}\equiv \delta_0 +\theta \delta_1
-\frac{\partial}{\partial\theta}
\label{1.12}
\end{equation}
\noindent

and the nilpotency condition gives the algebraic rules
\begin{eqnarray}
\delta_1 = \delta_{0}^2\nonumber \\
\delta_1 \delta_{0}=\delta_{0} \delta_1
\label{1.13}
\end{eqnarray}

Our purpose will be to extend to the quantum level the symmetry
described by (\ref{1.11}); according to the Q.A.P. \cite{11}
 for an arbitrary choice of the effective Lagrangian, the obstruction in
minimal order is given by:
\begin{equation}
\delta \Gamma= \Delta
\label{1.14}
\end{equation}
\noindent
where $\Delta$ is a local functional decomposed into its external fields
and $\theta$ dependent part as:
\begin{eqnarray}
\Delta&\equiv&\int d^4x \biggl[\Delta_0(x)+\theta\Delta_1(x)\biggr] \nonumber
\\
&+&\int d^4x \biggl[\gamma^{a\mu\nu}(x)
(\Theta_{\mu\nu 0}^a(x)+\theta\Theta_{\mu\nu 1}^a(x))
+\zeta^{a\rho} (x) (\Psi_{\rho 0}^a (x) \nonumber \\
&+&\theta\Psi_{\rho 1}^a (x))\biggr]
\label{1.15}
\end{eqnarray}

The renormalization program will be completed if the obstruction $\Delta$
in (\ref{1.15}) can be compensated by means of local counterterms
to be introduced in the effective Lagrangian.
Recall that the definition of the classical model allows $\theta$ dependent
 countertems only in that part of the action which contains external fields,
for this reason we decompose:
\begin{equation}
L_{eff}=L_{eff}^0 +\gamma^{a\mu\nu}(x)
(\hat{\Theta}_{\mu\nu 0}^a(x)+\theta\hat{\Theta}_{\mu\nu 1}^a(x))+
\zeta^{a\rho} (x) \hat{\Psi}_{\rho 0}^a (x)
\label{1.16}
\end{equation}
\noindent

and as subsidiary condition, we impose:
\begin{equation}
\frac{\partial L_{eff}^0}{\partial\theta}=0
\label{1.17}
\end{equation}

The power counting, Gparity and $\Phi \Pi$ Charge conservations
control the quantum numbers of the anomaly $\Delta$ and yield
contraints to the renormalization procedure; for this reason we shall summarize
in the following table all the global properties of the constituent fields
of the model:

\begin{tabular}{||l|l|l|l|}\hline\hline
Fields and parameters    & Dimension    & $\Phi \Pi$ Charge & Gparity \\
\hline
        g         &     -1         &    0              & \\  \hline
$B_{\mu\nu}^i(x)$ &      1       &      0            & - \\  \hline
$A_{\mu}^i(x)$    &      2       &        0          & + \\  \hline
$D_{\mu}^{ij}(x)$    &      1       &        0          &+ \\  \hline
$F_{\mu\nu}^i(x)$ &      3       &      0            &+ \\  \hline
$C_{\sigma}^i(x)$ &      1        & 1                &- \\   \hline
$C^i (x)$         &      1        & 2                 &- \\  \hline
$\bar{c}_{\sigma}^i(x)$ &      1        & -1           &-      \\   \hline
$\bar{c}^i (x)$         &      1        & -2           & -\\  \hline
$b_{\sigma}^i(x)$ &      2        & 0                &- \\   \hline
$\theta$         &      -1        & -1                &+  \\  \hline
$\gamma_i^{\mu\nu}(x)$&  2         & -1                  & -    \\ \hline
$\zeta_i^{\rho}(x)$&     2      & -2                   & -    \\ \hline
$\hat{\Theta}_{\mu\nu 0}^i(x)$&  2         &  1                 & -    \\
\hline
$\hat{\Theta}_{\mu\nu 1}^i(x)$&   3        &  2                 & -   \\ \hline
$\hat{\Psi}_{\rho 0}^i$&    2       &   2                & -   \\ \hline
$\Delta_0(x)$&   5        &      1             &   + \\ \hline
$\Delta_1(x)$&    6       &      2        &     +  \\ \hline
$\Theta_{\mu\nu 0}^i(x)$&  3         &      2             & -   \\ \hline
$\Theta_{\mu\nu 1}^i(x)$&   4        &       3            &  -  \\ \hline
$\Psi_{\rho 0}^i (x)$&     3     &          3         & -   \\ \hline
$\Psi_{\rho 1}^i (x)$&      4     &          4         &  -  \\ \hline
\end{tabular}

Now the external fields anomalies
are absent since we shall see that the consistency (cocycle) condition
indeed coincides with compensability (cobordism)  so
this part of the cohomology will be empty from the very definition.

In fact the cocycle condition:
\begin{equation}
\delta_{B.R.S.}\Delta=0
\label{1.18}
\end{equation}
\noindent
requires:
\begin{eqnarray}
\Psi_{\rho 1}^i (x)&=&\delta_0 \Psi_{\rho 0}^i (x) \nonumber \\
\Theta_{\mu\nu 1}^i(x)&=&\delta_0\Theta_{\mu\nu 0}^i(x))
\label{1.19}
\end{eqnarray}

On the other hand the compensability condition
\begin{equation}
\Delta=\delta_{B.R.S.}\hat\Delta
\label{1.20}
\end{equation}
\noindent

written in terms of the quantities defined in (\ref{1.16}) imposes
\begin{eqnarray}
\Psi_{\rho 1}^a (x)&=&\delta_1 \hat{\Psi}_{\rho }^a (x) \nonumber \\
\Psi_{\rho 0}^a (x)&=&\delta_0 \hat{\Psi}_{\rho }^a (x) \nonumber \\
\Theta_{\mu\nu 0}^a(x)&=&\delta_0\hat{\Theta}_{\mu\nu 0}^a(x)-\hat{\Theta}_
{\mu\nu 1}^a(x) \nonumber \\
\Theta_{\mu\nu 1}^a(x)&=&\delta_1\hat{\Theta}_{\mu\nu 0}^a(x)-
\delta_0\hat{\Theta}_{\mu\nu 1}^a(x)
\label{1.21}
\end{eqnarray}

It is straightforward to check, eliminating the hatted fields in (\ref{1.21})
by means of the relations (\ref{1.13}), that we reconstruct the
 cocycle conditions: therefore this cohomology subspace is empty.

We consider now the external field independent part of the anomaly
which is analyzed according to the $\theta$ dependence as
\begin{eqnarray}
\int d^4x\Delta_1(x) &=& \delta_1\Gamma \nonumber \\
\int d^4x\Delta_0(x) &=&\delta_0\Gamma
\label{1.22}
\end{eqnarray}

The cocycle (consistency) condition (\ref{1.17}) implies:

\begin{eqnarray}
\int d^4x\Delta_1(x)&=&\delta_0\int d^4x\Delta_0(x) \nonumber  \\
\delta_0\int d^4x\Delta_1(x)&=&\delta_1\int d^4x\Delta_0(x)
\label{1.23}
\end{eqnarray}
\noindent

Decomposing the anomalies according to their $\Phi.\Pi.$ charge content
 we obtain:
\begin{eqnarray}
\int d^4x\Delta_1(x)&=&\int d^4x\biggl[c^a(x)\Delta_{1,a}(x)+\sum_{m}
c_{\mu}^a(x){{\partial}_{\lambda_m}}^m c_{\nu}^b(x)
\Delta_{1,a,b,{\lambda}_m}^{\mu\nu}(x)\biggr] \nonumber \\
\int d^4x\Delta_0(x)&=&\int d^4x\biggl[c_{\mu}^a(x)
\Delta_{0,a}^{\mu}(x)\biggr]
\label{1.24}
\end{eqnarray}
\noindent

where $\Delta_{0,a}^{\mu}(x)$, $ \Delta_{1,a}(x)$ and
$\Delta_{1,a,b,{\lambda}_m}^{\mu\nu}(x)$ are $\Phi.\Pi.$ neutral.

The first cocycle condition(\ref{1.23}) furthermore gives:
\begin{eqnarray}
&\int& d^4x\biggl[c^a(x)\Delta_{1,a}(x)+\sum_{m}
c_{\mu}^a(x){{\partial}_{\lambda_m}}^m c_{\nu}^b(x)
\Delta_{1,a,b,{\lambda}_m}^{\mu\nu}(x)\biggr] \nonumber \\
&=&-\int d^4x\biggl[c^a(x){\biggl(D_{\mu a}^b \Delta_{0 b}^{\mu}(x)\biggr)}+
c_{\mu}^a(x)\delta_0\Delta_{0,a}^{\mu}(x)\biggr]
\label{1.25}
\end{eqnarray}
\noindent

and consequently the relations:
\begin{eqnarray}
\Delta_{1,a}(x)&=&-{\biggl(D_{\mu a}^b\Delta_{0 b}^{\mu}(x)\biggr)}
\sum_{m}{-1}^m
\biggl({{\partial}_{\lambda_m}}^m\Delta_{1,a,b,{\lambda}_m}^{\mu\nu}(x)
-(\mu\rightarrow\nu)(i\rightarrow j)\biggr) \nonumber \\
&=&D_\rho \frac{\delta}{\delta B_{\rho\nu}^b (x)}\int d^4y\Delta_{0,a}^{\mu}(y)
-\int d^4yD^{y}_\rho \frac{\delta}{\delta B_{\rho\nu}^b (y)}
\Delta_{0,a}^{\nu}(x)
\label{1.26}
\end{eqnarray}
\noindent

and the $c^a(x)$ content of $\Delta_{1,a}(x)$ disappears if and only if
${\biggl(D_{\mu a}^b \Delta_{0 b}^{\mu}(x)\biggr)}=0$.

This step allows us to considerably simplify our treatment in terms of
 local quantities; indeed the model is defined in a vector space whose
 entities are the monomials of local constituent fields and
their space-time derivatives considered as independent.

It is  useful to introduce in this space a new coordinate system, for the
uncharged $\Phi.\Pi.$ sector, such that each function of the fields and their
space-time derivatives is reparametrized in a different way.
If we grade this space in terms of  ordinary space-time derivatives,it is
a trivial observation that a monomial containing a space-time derivative
of order n can be written in terms of one containing  covariant derivatives
of the same order plus  terms of lower order in the covariant derivatives .

In this naive decomposition we have to take into account that while
in each monomial written in terms of the  ordinary space-time derivatives
, the derivatives tensorial content is by construction completely
symmetrized,  this is not so for the same quantities written in terms
of covariant derivatives.
Thus, to implement this symmetrization, we are compelled to
introduce the curvature tensor $gF_{\mu\nu}^j(x)$  and  in general the
symmetrized covariant derivatives of $gF_{\mu\nu}^j(x)$)
as independent fields .

This implies a one-to-one map between the space of monomials of
ordinary space-time derivatives of fields and that given by
symmetrized covariant derivatives (which we shall denote by $\{D_{\alpha}\}_n$)
of the same fields  plus
symmetrized covariant derivatives of $gF_{\mu\nu}^j(x)$, considered as
independent fields.

This notation has the further advantage : the coupling costant g (which has a
negative dimension and so its presence may cause problems with a lower limit
in the power counting analysis) is contained explicitely at the classical
level in the covariant derivatives  and in the curvature tensor
$gF_{\mu\nu}^j(x)$; in the new basis the coupling constant will disappear
from the B.R.S. operator, while in the action it is present at the tree level
as an inverse power $\frac{1}{g}$ (which has positive power counting
dimension).

In this framework, defining
\begin{equation}
L^a(x)= f^{abc}\sum_{n}
\{D_{\alpha}\}_n gF_{\mu\nu}^b(x)
 \frac{\delta}{\delta\{D_{\alpha}\}_nB_{\mu\nu}^c}
\label{1.27}
\end{equation}
\noindent

the Q.A.P. gives for $\Delta_{1,a}(x)$
\begin{equation}
L^a(x)\Gamma=
\Delta_{1,a}(x)
\label{1.28}
\end{equation}
\noindent

and  $\Delta_{1,a}(x)$ has dimension 5 and odd Gparity .

Defining:
\begin{equation}
L^{a\dag}(x)= f^{abc}\sum_{n}\{D_{\alpha}\}_n B_
{\mu\nu}^c\frac{\delta}
{\delta\{D_{\alpha}\}_n gF_{\mu\nu}^b (x)}
\label{1.29}
\end{equation}
\noindent

we compute the commutators:
\begin{eqnarray}
 \biggl[L^a(x),L^{a\dag}(x)\biggr]_{-} &=& \sum_n\{D_{\alpha}\}_n
gF_{\rho\sigma}^b(x)
 \frac{\delta}{\{D_{\alpha} \}_n gF_{\rho\sigma}^b(x)} \nonumber \\
 &-&\sum_n\{D_{\alpha}\}_nB_{\rho\sigma}^b(x)
 \frac{\delta}{\{D_{\alpha}\}_n B_{\rho\sigma}^b(x)}
=(N_{gF}-N_B) \equiv  S(x) \nonumber \\
 \biggl[L^a(x), S(x)\biggr]_{-}&=& -2L^a(x)
\label{1.30}
\end{eqnarray}
\noindent

The above relations yield the identitis:
\begin{eqnarray}
 \biggl[L^a(x), S(x)\biggr]_{-}\Gamma&=& -2L^a(x)\Gamma \nonumber \\
(N_{gF}-N_B-2)\Delta_{1,a}(x)&=&L^a(x)S(x)\Gamma
\label{1.31}
\end{eqnarray}
\noindent

and each monomial $\Delta_{1,a}(x)$ such that
\begin{equation}
(N_{gF}-N_B-2)\Delta_{1,a}(x)\neq 0
\label{1.32}
\end{equation}
\noindent

clearly admits a counterterm compensation.

After this step we are left with the possible breakings satisfying
\begin{equation}
(N_{gF}-N_B-2)\Delta_{1,a}(x)=0
\label{1.33}
\end{equation}
\noindent

which can be parametrized as
\begin{equation}
\Delta_{1,a}(x)=\Lambda_{abc0} S^{\mu\nu\rho\sigma\lambda}D_\lambda
F_{\mu\nu}^b(x)F_{\rho\sigma}^c(x)+\frac{1}{g}\Lambda_{abc1}
S_{\mu\nu\rho\sigma}gF_{\mu\nu}^b(x)gF_{\rho\sigma}^c(x)
\label{1.34}
\end{equation}
\noindent

where $\Lambda_{abc0}$,$\Lambda_{abc1}$,$S_{\mu\nu\rho\sigma\lambda}$
and $S_{\mu\nu\rho\sigma}$
are dimensionless group and Lorentz invariant tensors respectively.

Now $S_{\mu\nu\rho\sigma\lambda}$ vanishes since no dimensionless
 Lorentz invariant tensor with odd indices exists in four dimensions, and
the odd Gparity requirement forces
$S_{\mu\nu\rho\sigma}=\epsilon_{\mu\nu\rho\sigma}$
 $\Lambda_{abc}=d_{abc}$
where $d_{abc}$ is the symmetric  group invariant tensor, so we are left with:
\begin{equation}
\Delta_{1,a}(x)=a_{1}\frac{1}{g}d_{abc}
\epsilon_{\mu\nu\rho\sigma}gF_{\mu\nu}^b(x)
gF_{\rho\sigma}^c(x)
\label{1.35}
\end{equation}

Finally the expression in (\ref{1.35}) does not satisfy the
 consistency condition (\ref{1.26}) since:
\begin{eqnarray}
\Delta_{1,a}(x)&=&a_{1}D_{\mu d}^f \biggr[\frac{1}{g}d_{abc} f_{bfe}
\epsilon_{\mu\nu\rho\sigma}D_{\nu e}^d
gF_{\rho\sigma}^c(x)\biggr] \nonumber \\
&\neq& - {\biggl
(D_{\mu a}^b \Delta_{0 b}^{\mu}(x)\biggr)}
\label{1.36}
\end{eqnarray}
\noindent

and therefore the coefficient $a_1$ must vanish.

By (\ref{1.26}), the vanishing of $\Delta_{1,a}$ implies for
$\Delta^{\mu}_{0,a}$
the  constraints
\begin{equation}
{\biggl(D_{\mu a}^b \Delta_{0 b}^{\mu}(x)\biggr)}\equiv\partial_\mu
\Delta_{0 a}^{\mu}(x)-gf^{abc}A_{\mu}^c\Delta_{0 b}^{\mu}(x)= 0\\
\label{1.38}
\end{equation}
and:
\noindent
\begin{equation}
L^a(x)\Delta_{0 b}^{\mu}(x)=0\\
\label{1.39}
\end{equation}
\noindent
which have to be verified for $\Delta_{0 b}^{\mu}(x)$ of
dimension 4 and even Gparity.

This last equation admits the solution:
\begin{equation}
\frac{\delta}{\delta\{D_{\alpha}\}_nB_{\mu\nu}^c}\Delta_{0 b}^{\rho}(x)=
\{D_{\alpha}\}_n gF_{\mu\nu}^c(x)X_{\rho}( A,gF)^{-}+ \{D_{\alpha}\}_n
\epsilon^{\mu\nu\lambda\eta}gF_{\lambda\eta}^c(x)X_{\rho}( A,gF)^{+}
\label{1.40}
\end{equation}
\noindent

where $X_{\rho}( A,gF)^{+}$ $X_{\rho}( A,gF)^{-}$ are arbitrary
analytic functions of $( A,gF)$ of even$(+)$ or odd$(-)$ G-parity.
Dimensional analysis points to $n=0$ as the only possibility
, but in this case the $X_{\rho}$ functions have dimension one,
which is not satisfiable with their field content.

This argument excludes the anomalies:

\begin{equation}
\Delta_1=\int d^4x\biggl[\sum_{m}
c_{\mu}^a(x){{\partial}_{{\lambda}_m}} c_{\nu}^b(x)
\Delta_{1,a,b,{\lambda}_m}^{\mu\nu}(x)\biggr]
\label{1.41}
\end{equation}
\noindent
and we remain with the case
when $\Delta_{0 b}^{\mu}(x)$, considered as formal power series in 1/g,
is $B^a_{\mu\nu}$ independent and depends only on the $A^{a}_{\mu}$,
$gF_{\mu\nu}^c(x)$ fields and their covariant derivatives.
Dimensional analysis, G-parity and Lorentz index
content easily imply that it is impossile to construct a
$\Delta_{0 b}^{\mu}(x)$ which obeys (\ref{1.38}),
therefore no anomaly can occur in this model.

\section{Conclusions}
We have completely characterized the antisymmetric tensor field models in
the two regimes; in particular for the "Quantum extension 1" we have proved
that the integrated cohomology space of the corresponding B.R.S. operator is
empty both in the  neutral $\Phi\Pi$ sector and in the anomaly sector.
The first property, characteristic of the topological theories, here is not
sufficient to insure finiteness due to the lack of the vectorial supersymmetry
 and instead we find that there are two renormalization constants.

Concerning the "Quantum extension 2" where the analysis performed by means
of the $\theta$ parameter offers an alternative to the better known
 "higher powers in the external fields' technique, we also find that there are
no anomalies, thus clarifying the possibility of maintaining to all
perturbative
 orders the $\sigma$ model interpretation .

\section{Appendix A}

We report here the propagators of the gauge fields since they differ, due
to the presence of the $\intx F^{a\mu\nu} F^{a}_{\mu\nu}$, from the usual
case: selecting the bilinear part of the classical action (\ref{icl}) and
considering only the terms involving the fields $A^{a}_{\mu}(p),
B^{a}_{\rho\sigma}(p), b^{a}(p), b^{a}_{\lambda}(p), e^{a}(p)$ we have
to compute the inverse of the matrix,with rows and columns ordered accordingly,

\begin{equation}
\left( \matrix{-(lp^{2}+k)\delta_{\mu\nu}+lp_{\mu}p_{\nu}&i
\varepsilon_{\rho\sigma\tau\mu}p^{\tau}&-i
p_{\mu}&0&0\cr -i\varepsilon_{\rho'\sigma'\tau'\nu}p^{\tau}&0&0&{i \over 2}
(\delta_{\rho'\lambda'}p_{\sigma'} - \delta_{\sigma'\lambda'}p_{\rho'})&0\cr
ip_{\nu}&0&0&0&0\cr
0&-{i \over 2}
(\delta_{\rho\lambda}p_{\sigma} - \delta_{\sigma\lambda}p_{\rho})&0&-{1\over4}
(lp^{2} + k)\delta_{\lambda\lambda'} +{l\over4}p_{\lambda} p_{\lambda'}&-i
p_{\lambda}\cr 0&0&0&ip_{\lambda'}&0} \right)
\label{aa1}
\end{equation}

\noindent
The inverse is given by
\begin{equation}
\left(\matrix{0&{i\over{2p^{2}}}\varepsilon_{\rho\sigma\tau\mu}p^{\tau}&-{i
\over p^{2}}p_{\mu}&0&0\cr -{i\over{2p^{2}}}
\varepsilon_{\rho'\sigma'\tau'\nu'}p^{\tau'}&{(lp^{2} + k)\over{4p^{2}}}(
\delta_{\rho\rho'}\delta_{\sigma\sigma'}-
\delta_{\rho\sigma'}\delta_{\sigma\rho'})&0&{i \over{p^{2}}}
(\delta_{\rho'\lambda'}p_{\sigma'} - \delta_{\sigma'\lambda'}p_{\rho'})&0\cr
{i \over p^{2}}p_{\nu'}&0&{k\over{p^{2}}}&0&0\cr
0&-{i \over{p^{2}}}
(\delta_{\rho\lambda}p_{\sigma} - \delta_{\sigma\lambda}p_{\rho})&0&0&-{
i\over{p^{2}}}p_{\lambda}\cr
0&0&0&{i\over{p^{2}}}p_{\lambda'}&{k\over{4p^{2}}}}\right)
\label{aa2}
\end{equation}

\noindent
{}From the above we obtain the U.V. dimensions of the  gauge fields reported
in text as the canonical dimensions and the I.R. dimensions
\begin{center}
\begin{tabular}{|l|r|r|r|r|r|}\hline
&$A_{\mu}$ & $B_{\mu\nu}$ & $b$ & $b_{\mu}$ & $e$ \\ \hline
I.R.dim&2&1&1&2&1 \\ \hline
\end{tabular}
\end{center}

\noindent
It is also clear that the propagators in (\ref{aa2}) are analytic in the
$k$ parameter

\section{Appendix B}

We briefly consider the renormalization of the nilpotency condition
\begin{equation}
{\widehat D}\Gamma^{(n)}\equiv \intx ( {\partial_{\mu}\theta^{a}}{\delta \over
{\delta {{\widehat\gamma}^{a}_{\mu}}}} +{\partial_{\mu}A^{a\mu}}{\delta \over
{\delta {\xi}^{a}}})\Gamma^{(n)} = 0
\label{b1}
\end{equation}
\noindent
via the Q.A.P  at the first non-trivial order we have
\begin{equation}
{\widehat D}\Gamma^{(n)} = X^{(n)} + O(\hbar^{(n+1)})
\label{b2}
\end{equation}

\noindent
where $X^{(n)}$ is a local functional of canonical dimension $\leq 2$,
carrying two units of $\Phi\Pi$ charge and respecting the constraints
(\ref{213}). Accordingly it can be parametrized as
\begin{equation}
X^{(n)} = \intx \LP L^{a[bc]}{\partial_{\mu}A^{a\mu}}\theta^{b}\theta^{c}
 + R^{a(bc)}A^{a\mu}\theta^{b}{\partial_{\mu}\theta^{c}}\RP
\label{b3}
\end{equation}

\noindent
and the square (round) brackets denote symmetrization (antisymmetrization).
We easily find that
\begin{equation}
X^{(n)} = {\widehat D}{\widehat X^{(n)}}
\label{b4}
\end{equation}

\noindent
with
\begin{equation}
{\widehat X^{(n)}} = \intx \LP L^{a[bc]}\xi^{a}\theta^{b}\theta^{c}
 + R^{a(bc)}A^{a\mu}\theta^{b}{\widehat \gamma}^{c}_{\mu}\RP
\label{b5}
\end{equation}

\noindent
which amounts to the compensability condition of the breaking $X^{(n)}$.

\section{Appendix C}
In order to identify the local cohomology of the ${\widehat S}_{L}$ operator
 we  filter with the counting operator \cite{12}
\begin{equation}
\em N = \intx \LP A^{a\mu}{\delta \over {\delta{A^{a\mu}}}} +
{\widehat \gamma}^{a\mu\nu}
{\delta \over {\delta{{\widehat \gamma}^{a\mu\nu}}}}
+ \theta^{a}{\delta \over {\delta\theta^{a}}}+
\eta^{a}{\delta \over {\delta\eta^{a}}} +
{\widehat \Omega}^{a\mu}{\delta \over {\delta{{\widehat \Omega}^{a\mu}}}} \RP
\label{41}
\end{equation}

\noindent
and accordingly we have
\begin{equation}
{{\widehat S}_{L}} = S^{(0)} + S^{(1)} + S^{(2)} + S^{(3)}
\label{42}
\end{equation}

\noindent
which obey
\begin{eqnarray}
(S^{(0)})^{2} &=& 0 \nonumber \\
\{ S^{(0)}, S^{(1)}\} &=& -kD \nonumber \\
\{ S^{(0)}, S^{(2)}\} + (S^{(1)})^{2} &=& 0 \nonumber \\
\{ S^{(0)}, S^{(3)}\} + \{ S^{(1)}, S^{(2)}\} &=& 0 \nonumber \\
(S^{(2)})^{2} + \{ S^{(1)}, S^{(3)}\} &=& 0 \nonumber \\
\{ S^{(2)}, S^{(3)}\} &=& 0 \nonumber \\
(S^{(3)})^{2} &=& 0
\label{43}
\end{eqnarray}

\noindent
and $S^{(0)}$ is explicitely given by
\begin{eqnarray}
S^{(0)} &=&
\intx \LP {\partial_{\mu}\theta^{a}}{\delta \over {\delta{A^{a\mu}}}} +
\varepsilon^{\mu\nu\rho\sigma} \partial_{\nu} B_{\rho\sigma}
{\delta \over {\delta{\widehat \gamma}^{a\mu}}}
+ 2{\partial_{\mu} c^{a\nu}}{\delta \over {\delta{B^{a\mu\nu}}}} -
{\varepsilon^{\mu\nu\rho\sigma} \partial_{\rho}A^{a}_{\sigma}}
{\delta \over {\delta{\widehat \gamma}^{a\mu\nu}}} \nonumber \\
 &-& 2{\partial^{\nu}{\widehat \gamma}^{a}_{\mu\nu}}
{\delta\over {\delta{\widehat \Omega^{a}_{\mu}}}}
+{\partial_{\mu}{\widehat\gamma}^{a\mu}}{\delta \over {\delta{\xi^{a}}}} -
{\partial_{\mu}{\widehat \Omega}^{a\mu}}{\delta \over {\delta{\eta^{a}}}} +
{\partial_{\mu}d^{a}}{\delta \over {\delta{c_{\mu}^{a}}}} \RP
\label{44}
\end{eqnarray}

When operating on functions, $S^{(0)}$ can be rewritten as an ordinary
differential operator; setting
\begin{eqnarray}
A^{+a}_{\mu,\nu} &=& {1 \over 2} (\partial_{\mu}A^{a}_{\nu}
+ \partial_{\nu}A^{a}_{\mu}) \nonumber \\
A^{-a}_{\mu,\nu} &=& {1 \over 2} (\partial_{\mu}A^{a}_{\nu}
- \partial_{\nu}A^{a}_{\mu}) \nonumber \\
c^{+a}_{\mu,\nu} &=& {1 \over 2} (\partial_{\mu}c^{a}_{\nu}
+ \partial_{\nu}c^{a}_{\mu}) \nonumber \\
c^{-a}_{\mu,\nu} &=& {1 \over 2} (\partial_{\mu}c^{a}_{\nu}
- \partial_{\nu}c^{a}_{\mu}) \nonumber \\
B^{-a}_{\mu\nu,\rho} &=& {1 \over 3}(\partial_{\rho}B^{a}_{\mu\nu} +
\partial_{\mu}B^{a}_{\nu\rho} + \partial_{\nu}B^{a}_{\rho\mu}) \nonumber \\
B^{+a}_{\mu\nu,\rho} &=& {1 \over 3}(2\partial_{\rho}B^{a}_{\mu\nu} -
\partial_{\mu}B^{a}_{\nu\rho} - \partial_{\nu}B^{a}_{\rho\mu})
\label{bb2}
\end{eqnarray}

\noindent
and denoting the space-time partial derivatives of the fields with an index
after a comma, we have
\begin{eqnarray}
S^{(0)} &=& \theta^{a,\mu}{\partial \over{\partial A^{a\mu}}} +
 \theta^{a,\mu\nu}{\partial \over{\partial A^{+a\mu,\nu}}} +
 \theta^{a,\mu\nu\rho}{\partial \over{\partial A^{+a\mu,\nu\rho}}} \nonumber \\
&+& \varepsilon_{\mu\nu\rho\sigma}(B^{-a\nu\rho,\sigma}
{\partial \over{\partial {\widehat\gamma}^{a\mu}}} +
B^{-a\nu\rho,\sigma\tau}
{\partial \over{\partial {\widehat\gamma}^{a\mu,\tau}}}) \nonumber \\
&+& 4(c^{-a\mu,\nu}{\partial \over{\partial B^{a\mu\nu}}} +
c^{-a\mu,\nu\rho}{\partial \over{\partial B^{+a\mu\nu,\rho}}} +
c^{-a\mu,\nu\rho\sigma}
{\partial \over{\partial B^{+a\mu\nu,\rho\sigma}}}) \nonumber \\
&-&2\varepsilon^{\mu\nu\rho\sigma}(A^{-a}_{\rho,\sigma}
{\partial \over{\partial {\widehat\gamma}^{a\mu\nu}}} +
A^{-a}_{\rho,\sigma\tau}
{\partial \over{\partial {\widehat\gamma}^{a\mu\nu,\tau}}} +
A^{-a}_{\rho,\sigma\tau\lambda}
{\partial \over{\partial {\widehat\gamma}^{a\mu\nu,\tau\lambda}}}) \nonumber \\
&+& {\widehat\gamma}^{a\mu}_{\mu}{\partial \over{\partial \xi^{a}}} -
{\widehat\Omega}^{a\mu}_{\mu}{\partial \over{\partial \eta^{a}}} +
2({\widehat\gamma}^{a\mu\nu}_{\nu}
{\partial \over{\partial {\widehat\Omega}^{a\mu}}} +
{\widehat\gamma}^{a\mu\nu,\tau}_{\nu}
{\partial \over{\partial {\widehat\Omega}^{a\mu,\tau}}}) \nonumber \\
&+&\partial d^{a,\mu}{\partial \over{\partial c^{a\mu}}} +
 d^{a,\mu\nu}{\partial \over{\partial c^{+a\mu,\nu}}} +
 d^{a,\mu\nu\rho}{\partial \over{\partial c^{+a\mu,\nu\rho}}} +
 d^{a,\mu\nu\rho\sigma}{\partial \over{\partial c^{+a\mu,\nu\rho\sigma}}}
\label{bb3}
\end{eqnarray}

\noindent
In the above expression all indices after the comma are completely
simmetrized and we have taken into account that $S^{(0)}$ acts on functions
of canonical dimension $\leq 4$. In this space we can define $S^{(0)\dagger}$
as
\begin{eqnarray}
S^{(0)\dagger} &=& A^{a\mu}{\partial \over{\partial\theta^{a,\mu}}} +
 A^{+a\mu,\nu}{\partial \over{\partial \theta^{a,\mu\nu}}} +
A^{+a\mu,\nu\rho}{\partial \over{\partial \theta^{a,\mu\nu\rho}}} \nonumber \\
&+& \varepsilon_{\mu\nu\rho\sigma}({\widehat\gamma}^{a\mu}
{\partial \over{\partial B^{-a\nu\rho,\sigma}}} +
{\widehat\gamma}^{a\mu,\tau}
{\partial \over{\partial B^{-a\nu\rho,\sigma\tau}}}) \nonumber \\
&+& 4(B^{a\mu\nu}{\partial \over{\partial c^{-a\mu,\nu}}} +
B^{+a\mu\nu,\rho}{\partial \over{\partial c^{-a\mu,\nu\rho}}} +
B^{+a\mu\nu,\rho\sigma}
{\partial \over{\partial c^{-a\mu,\nu\rho\sigma}}}) \nonumber \\
&-&2\varepsilon_{\mu\nu\rho\sigma}({\widehat\gamma}^{a\mu\nu}
{\partial \over{\partial A^{-a}_{\rho,\sigma}}} +
{\widehat\gamma}^{a\mu\nu,\tau}
{\partial \over{\partial A^{-a}_{\rho,\sigma\tau}}} +
{\widehat\gamma}^{a\mu\nu,\tau\lambda}
{\partial \over{\partial A^{-a}_{\rho,\sigma\tau\lambda}}}) \nonumber \\
&+& { \xi^{a}}{\partial \over \partial{\widehat\gamma}^{a\mu}_{\mu}} -
\eta^{a}{\partial \over{\partial {\widehat\Omega}^{a\mu}_{\mu}}} +
2({\widehat\Omega}^{a\mu}
{\partial \over{\partial {\widehat\gamma}^{a\mu\nu}_{\nu}}} +
{\widehat\Omega}^{a\mu,\tau}
{\partial \over{\partial {\widehat\gamma}^{a\mu\nu,\tau}_{\nu}}}) \nonumber \\
&+& c^{a\mu}{\partial \over{\partial d^{a,\mu}}} +
c^{+a\mu,\nu}{\partial \over{\partial d^{a,\mu\nu}}} +
c^{+a\mu,\nu\rho}{\partial \over{\partial d^{a,\mu\nu\rho}}} +
 c^{+a\mu,\nu\rho\sigma}{\partial \over{\partial d^{a,\mu\nu\rho\sigma}}}
\label{bb4}
\end{eqnarray}

Now the local cohomology of $S^{(0)}$ is identified by $X^{ch}(x)$,
solution of
\begin{equation}
\{ S^{(0)}, S^{(0)\dagger}\}X^{ch}(x) = 0
\label{bb5}
\end{equation}

\noindent
By direct computation $\{ S^{(0)}, S^{(0)\dagger}\}$ is an operator
which counts all fields except $\theta^{a},d^{a}$ without space time
derivatives;  therefore we find
\begin{equation}
X^{ch}(x) \equiv X^{ch}(\theta^{a}(x),d^{a}(x))
\label{bb6}
\end{equation}

\noindent
The above result must be further restricted by the condition
\begin{equation}
N^{a}X^{ch}(x) = 0
\label{bb7}
\end{equation}

\noindent
 since all other constraints are already satisfied; this yields
\begin{equation}
X^{ch}(x) \equiv X^{ch}(\theta^{a}(x))
\label{bb8}
\end{equation}

Let us now return to the local cohomology of ${\widehat S}_{L}$; setting
\begin{equation}
X(x) = \sum_{n=1}X^{(n)}
\label{bb9}
\end{equation}

\noindent
we obtain for $X^{(n)}$ the recurrence
\begin{equation}
S^{(0)}X^{(n)}(x) + S^{(1)}X^{(n-1)}(x) + S^{(2)}X^{(n-2)}(x) +
S^{(3)}X^{(n-3)}(x) = 0
\label{bb10}
\end{equation}

The solution of (\ref{bb10}) is easily found if we observe that
\begin{eqnarray}
S^{(1)}X^{ch}(x) &=& {1 \over 2}f^{abc}\theta^{b}\theta^{c}{\partial \over
{\partial\theta^{a}}}X^{ch}(x) \nonumber \\
S^{(2)}X^{ch}(x) &=& 0 \nonumber \\
S^{(3)}X^{ch}(x) &=& 0
\label{bb11}
\end{eqnarray}

\noindent
and take into account the relations (\ref{43}).Let us analyze the first
few terms; for $n=1$ we have
\begin{equation}
S^{(0)}X^{(1)}(x) =0
\end{equation}

\noindent
with solution
\begin{equation}
X^{(1)}(x) = S^{(0)}{\widehat X}^{(1)}(x) + X^{ch(1)}(\theta)
\label{bb12}
\end{equation}

\noindent
which substituted into the next $(n=2)$ relation gives
\begin{equation}
S^{(0)}(X^{(2)}(x) -S^{(1)}{\widehat X}^{(1)}(x)) + S^{(1)}X^{ch(1)}(x) = 0
\end{equation}

\noindent
Now the last term on the l.h.s. of the above equation still belongs
to the cohomology of $S^{(0)}$ and therefore it must separetely vanish,
leading to the solution
\begin{eqnarray}
X^{(2)}(x) &=& S^{(0)}{\widehat X}^{(2)}(x) + S^{(1)}{\widehat X}^{(1)}(x) +
X^{ch(2)}(\theta) \nonumber \\
{1 \over 2}f^{abc}\theta^{b}\theta^{c}{\partial \over
{\partial\theta^{a}}}X^{ch(1)}(x) &=& 0
\label{bb13}
\end{eqnarray}

\noindent
The procedure iterates and we find
\begin{eqnarray}
S^{(0)}(X^{(n+3)}(x) &-& S^{(1)}{\widehat X}^{(n+2)}(x) -
S^{(2)}{\widehat X}^{(n+1)}(x) - S^{(3)}{\widehat X}^{(n)}(x)) = 0 \nonumber \\
S^{(1)}X^{ch(n+2)}(x) &=& 0
\label{bb14}
\end{eqnarray}

\noindent
which is the result given in the text.

Furthermore the general local functional ${\widehat\Gamma}^{0}$
of dimensions $\leq 4$, even G-parity, $\Phi\Pi$ charge
 $-1$ and obeying
\begin{eqnarray}
{\widehat D}{\widehat \Gamma}^{(0)} &=& 0\nonumber \\
G^{a}{\widehat \Gamma}^{(0)} &=& 0 \nonumber \\
{\widehat\Gamma}^{0}&\not=& {\widehat S}_{L}{\widehat{\widehat\Gamma}}^{0}
\label{c1}
\end{eqnarray}

\noindent

can be parametrized as:
\begin{eqnarray}
{\widehat \Gamma}^{(0)} &=&\intx \LP a_{1}{\widehat\Omega}^{a\mu} c^{a}_{\mu} +
a_{2}{\widehat\Omega}^{a\mu}\partial_{\mu} d^{a} +
a_{3}\varepsilon^{\mu\nu\rho\sigma}{\widehat \gamma}^{a}_{\mu\nu}
B{a}_{\rho\sigma} \nonumber \\
&+& a_{4} f^{abc}{\widehat \gamma}^{a}_{\mu\nu}
{\widehat \gamma}^{b\mu\nu}\theta^{c} +
a_{5}\varepsilon^{\mu\nu\rho\sigma}{\widehat \gamma}^{a}_{\mu\nu}
\partial_{\rho} A^{a\sigma} +
a_{6}f^{abc}\varepsilon^{\mu\nu\rho\sigma}{\widehat \gamma}^{a}_{\mu\nu}
A^{b}_{\rho}A^{c}_{\sigma} \nonumber \\
&+& a_{7}\xi^{a}\theta^{a} + a_{8}{\widehat\gamma}^{a\mu}A^{a}_{\mu} \RP
\label{c2}
\end{eqnarray}

\noindent
and from (\ref{c1}) we find $a_{1} = a_{2} = a_{3} = a_{7} = a_{8} = 0$.
The remaining three parameters are not independent since we still have
to consider ${\widehat\Gamma}^{0}$ modulo terms of the type
${\widehat S}_{L}{\widehat{\widehat\Gamma}}^{0}$ which do not contribute
to the counterterms. Now
\begin{equation}
{\widehat{\widehat\Gamma}}^{0} \propto \intx {\widehat \gamma}^{a}_{\mu\nu}
{\widehat \gamma}^{a\mu\nu}
\label{c3}
\end{equation}

\noindent

and we finally obtain that there are only two nonvanishing parameters,
for instance $a_{4}, a_5$ .


\begin{thebibliography}{99}
\bibitem{1} V. I. Ogievetsky and I. V. Polubarinov, Yad. Fiz. 4(1968)210 \\
            S. Deser, Phys.Rev178(1969)1931 \\
            K. Hayashi, Phys.Lett.B44(1973)497 \\
            J. Kalb and P. Ramond, Phys.Rev.D9(1974)2273 \\
            Y. Nambu, Phys.Reports23(1976)250
\bibitem{2} W.Siegel Phys. Lett. 93B (1980) 275 \\
            D. Z. Freedman, P. K. Townsend, Nucl.Phys.B177(1981)282 \\
            P.K.Townsend, Phys.Lett.B88(1979)97 \\
            L.Baulieu J.Thierry-Mieg Phys.Lett 144B(1983) 221 \\
            L.Baulieu J.Thierry-Mieg Nucl. Phys.B 228(1985) 259 \\
            F.S.Fradkin and A.A.Tseytlin, Ann.Phys.162(1985)31 \\
            S. P. de Alwis, M. T. Grisaru and L. Mezincescu,
            Phys.Lett.B190(1987)122 \\
            N.K.Nielsen, Nucl.Phys.B332(1990)391
\bibitem{3} S. P. de Alwis, M. T. Grisaru and L. Mezincescu,
            Nucl.Phys.B303(1988)57
\bibitem{4}T.E.Clark,C.H.Lee and S.T.Love, Nucl.Phys.B308(1988)379

\bibitem{5} G. T. Horowitz, Comm.Math.Phys.125(1989)417 \\
            G. T. Horowitz and M. Srednicki,Comm.Math.Phys.130(1990)83 \\
            M. Blau and G. Thompson, Ann.Phys.(N.Y.)205(1991)130 \\
            M. Blau and G. Thompson, Phys.Lett.B255(1991)535
\bibitem{6} E. Guadagnini, N. Maggiore and S. P. Sorella,
            Phys.Lett.B255(1991)65
\bibitem{7} N. Maggiore and S. P. Sorella, Nucl.Phys.B377(1992)236  \\
            N.Maggiore and S.P.Sorella preprint UGVA-DPT 1992/04-761
\bibitem{8} J. Thierry-Mieg and L. Baulieu, Nucl.Phys.B228(1983)259 \\
             A. H. Diaz, Phs.lett.B203(1988)408 \\
             J. Thierry-Mieg, Nucl.Phys.B335(1990)334
\bibitem{9} A. Blasi, O. Piguet and S.P. Sorella,
                        Nucl.Phys.B356(1991)154
\bibitem{10}  J. Manes, R. Stora and B. Zumino, Comm.Math.phys.102(1985)157 \\
              B. Zumino, Nucl.Phys.B253(1985)477
\bibitem{11} Y.M.P Lam, Phys.Rev.D6(1972)2145, 2161 \\
             T.E. Clark and J.H. Lowenstein, Nucl.Phys.B113(1976)109
\bibitem{12} J.A. Dixon,  Cohomology and Renormalization of gauge
                theories, Imperial College preprint-1977 \\
             G. Bandelloni, J.Math.Phys.28(1987)2775
\bibitem{13} B.de Witt J.W. van Holten Phys.Lett.79B(1979) 389
\bibitem{14} G.Bandelloni J.Math.Phys 32(1991)3491


\end{thebibliography}
\end{document}